\documentclass[12pt]{article}
\usepackage{geometry} % see geometry.pdf on how to lay out the page. There's lots.
\usepackage[parfill]{parskip}
\usepackage{graphicx}
\usepackage{amssymb}

\title{Eliminating spurious poles from gauge-theoretic amplitudes}
\author{Andrew Hodges\thanks{ andrew.hodges@wadh.ox.ac.uk, http://www.twistordiagrams.org.uk}\\{\footnotesize  {\it Wadham College, University of Oxford, Oxford OX1 3PN, U.K.}}}

\date{10 May 2009} % delete this line to display the current date

\begin{document}

\maketitle

\begin{abstract}
This note addresses the problem of spurious poles in gauge-theoretic scattering amplitudes.  New twistor coordinates for the momenta are introduced, based on the concept of dual conformal invariance.  The cancellation of spurious poles for a class of NMHV amplitudes is greatly simplified in these coordinates. The poles are eliminated altogether by defining a new type of twistor integral, dual to twistor diagrams as previously studied, and considerably simpler. The geometric features indicate a supersymmetric extension of the formalism at least to all NMHV amplitudes, allowing the dihedral symmetry of the super-amplitude to be made manifest. More generally, the definition of `momentum-twistor' coordinates suggests a powerful new approach to the study of scattering amplitudes.

\end{abstract}

\section{The problem of spurious poles}
In this note we tackle the problem which plagues the representation of all gauge-theoretic scattering amplitudes except the MHV and $\overline{{\rm MHV}}$. The on-shell recursion relation due to Britto, Cachazo, Feng and Witten (2004, 2005), referred to as BCFW in what follows, has enormously simplified the calculation of amplitudes in terms of momentum-space spinors. But it leaves the results in the form of a sum of rational functions of momenta which, individually, have singularities which are not present in the amplitude itself. These are the spurious poles. 
We will start with the simplest example of this phenomenon.

\subsection{The split-helicity 6-field NMHV amplitude}

We consider the (colour-stripped) amplitude $A(1^- 2^- 3^- 4^+ 5^+ 6^+)$. This has a special simplicity because of the `split-helicity' distribution of $+$ and $-$ helicities into just two sets around the ring. 

The BCFW recursion can readily be applied to evaluate the amplitude, but the result obtained depends on the selection of the pivoting pair. Two quite different expressions result, one in two terms and one in three:

\begin{equation}
\left( \begin{array}{lr}
&\displaystyle{\frac{[4|5+6|1\rangle^3}{ [34][23]\langle56\rangle \langle61\rangle   [2|3+4|5\rangle S_{234}}}\\
\vspace{0mm}\\
 +&
\displaystyle{ \frac{[6|1+2|3\rangle^3}{  [61][12]\langle34\rangle \langle45 \rangle [2|3+4|5\rangle S_{612}}}
\end{array}
\right)\: \delta(\sum_{i=1}^{6}p_i)
\label{eqn:first}
\end{equation}
and
\begin{equation}
\left(
\begin{array}{lr}
 &\displaystyle{\frac{(S_{123})^3}{[12][23]\langle45\rangle \langle56\rangle [1|2+3|4\rangle  [3|4+5|6\rangle}}\\
 \vspace{0mm}\\
+&
\displaystyle{\frac{\langle12\rangle^3[45]^3}{\langle 16\rangle [34]  [3|4+5|6\rangle [5|6+1|2\rangle S_{612}}}\\
\vspace{0mm}\\
+&
\displaystyle{\frac{\langle23\rangle^3[56]^3}{\langle 34\rangle [16]  [1|2+3|4\rangle [5|6+1|2\rangle S_{234}}}
\end{array}
\right)\: \delta(\sum_{i=1}^{6}p_i)\,.
\label{eqn:second}
\end{equation}

It is far from obvious that these two representations of the amplitude are equivalent. Their equality is guaranteed by the derivation of BCFW recursion from the Feynman rules, but is difficult to show directly. Nor is it immediately apparent that the poles $ [2|3+4|5\rangle, [1|2+3|4\rangle, [3|4+5|6\rangle, [5|6+1|2\rangle$ are spurious, each of them cancelling, apparently miraculously, when the terms are added. (It does of course follow from the equivalence of the expressions that these poles must be spurious, but this does not suggest a natural way of seeing the cancellations.) Taking another approach, it is straightforward to see from the Feynman expansion that singularities can only arise from the vanishing of the scalar invariants $S_{ij}, S_{ijk}$ etc., and this gives another proof that any poles other than these must be spurious. But again, such an argument depends on a lengthy sequence of algebraic manipulations, involving the very gauge-dependent terms that the BCFW recursion has so marvellously eliminated.

It is not difficult to see, in general terms, how such spurious singularities can arise. A key ingredient of the BCFW recursion is the application of complex analysis on $\mathbb{CP}^1$ to produce an identity based on {\em partial fractions}. Such identities naturally give rise to spurious poles. As an analogy, consider:  
\begin{equation}
\frac{1}{(z-v)(z-w)} = \frac{1}{(v-w)} \frac{1}{(z-v)} + \frac{1}{(w-v)}\frac{1}{(z-w)}.
\label{eqn:zvw}
\end{equation} 
If  $v\ne w$, this can be established by considering the behaviour of each side as an analytic function of $z$. The agreement of the residues at $z=v, z=w$, and  the regularity at $z=\infty$, suffice. But the identity has a spurious pole at $v=w$, where the residue calculus formula fails to apply. The spurious poles in the amplitude likewise correspond to coincidences of parameters. For instance, the vanishing of $[1|2+3|4\rangle $ implies that $ S_{23}S_{1234} = S_{123}S_{234}.$ 

Algebraic verification of the equivalence of these expressions is difficult because in addition to use of the spinor identity $\epsilon_{AB}\epsilon_{CD} + \epsilon_{AC}\epsilon_{DB}+ \epsilon_{AD}\epsilon_{BC} = 0$ (sometimes called Schouten's identity), it calls on repeated applications of the four constraints imposed by the $\delta$-function.  Field theorists have generally chosen a spinor basis in which to express these constraints, then used computer-supported algebra.  

\subsection{Super-amplitudes and dihedral symmetry}

 It is now well understood that super-symmetric extension eliminates the restriction on helicities in the original BCFW recursion, and greatly increases its power.  Arkani-Hamed, Cachazo and Kaplan (2008) have given an extensive account of the significance of this development for physical theory. The concept of {\em super-amplitude} greatly simplifies scattering theory by treating all helicity cases at once. In particular, as a function of $n$ \hbox{(super-)}momenta $p_i^a$ the super-amplitude has a simple symmetry: it must be invariant under $i \rightarrow i+1$ and $i\rightarrow n-i$. 

Yet this simple {\em dihedral symmetry} is  hidden in the (super-)BCFW expansion. The difficulty of showing the equivalence of expressions (\ref{eqn:first}) and (\ref{eqn:second}) is just one aspect of this larger problem. Their equality is just one super-component of a  identity for six super-fields, asserting a hidden $D_6$ symmetry. This identity was identified as the {\em hexagon identity} of (Hodges 2005b),  which  presented an earlier version of super-BCFW recursion through the extension of the {\em twistor diagram} formalism. 

Arkani-Hamed et al.\ (2009) have recently given the analogue of this  formalism for split-signature theory, whilst Mason and Skinner (2009) have also given a parallel derivation. (The advantage of the split-signature case is that the requisite integrations can be stated precisely, without the difficulty of defining contours that pervades the Minkowski space theory. The disadvantage is that it lacks the physical content of scattering theory as a process connecting past and future.)  This confluence has greatly stimulated fresh interest in the problems of the hidden dihedral symmetry and the spurious poles, because  these {\em combinatorial} features are essentially the same in split-signature theory as in Minkowski space. 

The spurious poles are problematic from a practical computational point of view, as well as presenting a long-standing difficulty in the theory. For more than six fields, the complexity and asymmetry only increases. For the NMHV amplitude for $n$ fields, a BCFW recursion gives rise to $(n-3)(n-4)/2$ terms, each choice of pivots giving rise to a different sum of terms. (For special helicity configurations, as with the split-helicity six-field case, some of these terms vanish, giving a formula with fewer terms.) The general situation for tree-level amplitudes is even more complex.
For $n$ fields the amplitudes require a total of $\frac{(2n-6)!}{(n-3)!(n-2)!}$ terms.  (This was noted in Hodges (2005b), as an aspect of the twistor diagram form of the BCFW recursion;  Arkani-Hamed et al.\ (2009) make further interesting suggestions based on the Catalan number that appears in this combinatorial formula.) Within this, the sector with $n-r$ fields of negative helicity and $r$ of positive helicity accounts for $\frac{(n-4)!(n-3)! }{(r-2)!(n-r-1)!(r-1)!(n-r-2)!}$ terms. It is, of course, a considerable triumph that explicit formulas for these terms can now be given, based on carrying out the BCFW recursion for $n$ steps. But these statements of the answers are no less complex than the recursion relation itself.
 
These formulas conceal the $D_n$ dihedral symmetry of the $n$-field amplitudes. They also conceal the identities which follow from the $D_m$ dihedral symmetry of the $m$-field amplitudes, where $m < n$, which can be used to put the $n$-field amplitudes in a more symmetrical form.  One of the advantages of the  twistor diagram form of the super-BCFW expansion (Hodges 2005b, 2006), is that it gives a graphical form to these identities. This is also noted and exploited by Arkani-Hamed et al.\ (2009). 

 Explicitly, the `square identity', which expresses the $D_4$ symmetry of the $n=4$ super-amplitude, is needed to get even the manifest $D_3$ symmetry of the $n=6$ NMHV amplitudes. The $D_6$ `hexagon identity' can then be used to show the $D_4$ symmetry of the $n = 8$ NNMHV amplitudes (Hodges 2005b, 2006). The full $D_8$ symmetry of these amplitudes then resides in a formidable 40-term identity which has probably never been checked explicitly. Twistor diagrams help in locating, expressing and using these identities, but they have not, as yet, shown how to derive the identities without reference to the Feynman theory on which BCFW is based.

Not all authors have found it important to focus upon these hidden symmetries, even in  the most advanced recent treatments. Mason and Skinner (2009), for instance, simply present one particular $n = 8$ NNMHV summation as an application of their calculus, placing their emphasis on the super-conformal invariance of the individual terms, as elegantly derived from super-twistor geometry. But Arkani-Hamed has made a point of the idea that such summations fail to present important aspects of the solution. His emphasis on the significance of the spurious poles has stimulated the present enquiry. (He has also emphasised that the hidden part of the symmetry can also be seen as a {\em parity} symmetry --- invariance when the r\^{o}le of twistors and dual twistors is interchanged.) Can twistor geometry go further in bringing out what is hidden, and so not just translating BCFW  but adding new content to the description of amplitudes? Yes, it can.

\section{Dual conformal symmetry}

In what follows we shall generally be considering  sequences of $n$ spinors, vectors, or twistors with a cyclical property. We use throughout the natural convention in which the labelling is  modulo $n$, allowing the $n$th element of a sequence also to be called the 0th.

A first observation is that the equivalence of (\ref{eqn:first}) and (\ref{eqn:second}) should be seen as a five-term identity for 12 independent complex spinors, subject to four holomorphic constraints, as indicated by the $\delta$-function. Imposing a reality condition on the momentum, so that the spinors are complex conjugates, does not simplify the question. 

\subsection{Momentum-twistor coordinates}

Nothing in the expressions (\ref{eqn:first}), (\ref{eqn:second}), suggests an immediate connection with twistors.
Nevertheless, the key step is to introduce new twistor-valued coordinates with which to express the content of these momentum spinors. We shall call these {\em momentum-twistors}. 

Using  Penrose's original conventions  for classical twistor geometry, twistors $Z^{\alpha}$ are written as ($\omega^A, \pi_{A'})$, the spinors being referred to as $\omega$- and $\pi$-parts of the twistor. A point $x^a$  in complexified Minkowski space $\mathbb{CM}$ is said to lie on the $\alpha$-plane of the twistor $Z^{\alpha}$, if  $\omega^A = ix^{AA'}\pi_{A'}$.  Twistors with vanishing $\pi$-parts have no such finite  $x^a$, and  correspond to an $\alpha$-plane in the null cone at infinity in the compactification of $\mathbb{CM}$. If the skew two-index twistor  $P^{[\alpha} Q^{\beta]}$ does not vanish, it defines a projective line in twistor space; if $\pi_P^{A'}\pi_{QA' }\ne 0$  then this corresponds to a finite point in $\mathbb{CM}$. Dual twistor space is defined in the standard sense of vector and projective spaces; dual twistors have the form $(\pi_A, \omega^{A'})$ and are, projectively, planes in twistor space. (It is often convenient to use the projective terms point, line, and plane even when the twistors are given a scale.) The complex conjugate $\bar{Z}_{\alpha}$ is defined as an element in dual twistor space, and the resulting $(++--)$ pseudo-norm ensures that twistors represent the conformal group on (real, compactified) Minkowski space. It is remarkable, however, how much structure is purely holomorphic, and exists independently of this definition.  (That our momentum-spinor identities are purely holomorphic is an example of this.) In contrast, the mechanism for the breaking of conformal invariance, by picking out the null cone at infinity, is an all-pervasive aspect of the application of twistor theory to scattering amplitudes. It appears in twistor algebra through the special skew two-index twistors $I_{\alpha \beta}, I^{\alpha \beta},$ which represent the vertex of the null cone at infinity.

The $\pi$-part of a twistor is obtained by transvecting it with $I_{\alpha \beta}$, as $I_{\alpha \beta}Z^{\beta} = (0, \pi^{A'})$.  A useful formula relating twistor and space-time geometry is that if $P^{\alpha}, Q^{\alpha}$ define a line in projective twistor space, corresponding to a point $x^a$ in $\mathbb{CM}$, whilst $R^{\alpha},  S^{\alpha}$ similarly define $y^a$, then the displacement tensor
\begin{equation}D^{\alpha}_{\beta} =  \frac{I^{\alpha\kappa}\epsilon_{\kappa \lambda \mu \nu}P^{\lambda}Q^{\mu}R^{[\nu}S^{\sigma ]}I_{\sigma \beta}}{(I_{\lambda \mu}P^{\lambda}Q^{\mu})(I_{\nu\sigma}R^{\nu}S^{\sigma})}
\label{eqn:displace}
\end{equation}
has only one non-vanishing component, namely $(x-y)^{AB'}$. Furthermore
\begin{equation} (x-y)^2 = -2 \frac{\epsilon_{\lambda \mu \nu \sigma}P^{\lambda}Q^{\mu}R^{\nu}S^{\sigma }}{(I_{\lambda \mu}P^{\lambda}Q^{\mu})(I_{\nu\sigma}R^{\nu}S^{\sigma})} \, .
\end{equation}
 In choosing these momentum-twistor coordinates we are motivated by the concept of {\em region space} on which {\em dual conformal symmetry} is defined. This is an affine space in which the external null momenta $p_i^a$ are defined as differences $x_i^a - x_{i-1}^a$, so that momentum conservation is expressed by $x_0^a = x_{n}^a$. 

\begin{figure}[h] %  figure placement: here, top, bottom, or page
   \centering
   \includegraphics[width=195px]{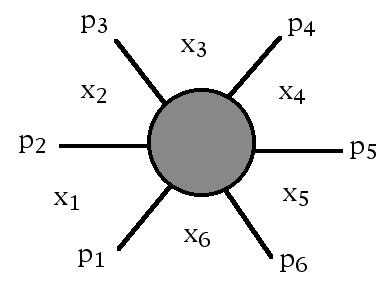} 
   \caption{Regions for a process with outgoing null momenta, with $p_i = x_i - x_{i-1}$.}
   \label{fig:ex1}
\end{figure}

The concept of region is only well-defined when the momenta have a sequential structure, such as arises from the colour-ordering. It is worth noting that dual conformal symmetry only holds for the colour-stripped amplitudes for the  planar Feynman graphs. Unlike the conformal symmetry in Minkowski space, it does not characterise the entire physical theory, but has a secondary character.
 
The concept of region space, and of conformal symmetry in this space, has long played an important role in the calculation of loop amplitudes for gauge theory. Recently, further inspiring discoveries have been made about dual super-conformal symmetry by Drummond et al., (2006),  Drummond,  Henn, Korchemsky and Sokatchev (2008),  Brandhuber et al.\ (2008), where, again, it is the extension to the full N=4 gauge theory with loops that is the driving force. In contrast, the discussion in this note is limited to tree-amplitudes, but the fact that this symmetry has such a powerful r\^{o}le in the full theory is part of the motivation. Note that Drummond and Henn (2008), in giving an explicit statement of all tree amplitudes, in a form which shows their dual conformal symmetry, still present the summation of many terms with spurious poles, and with all the hidden identities. Thus dual conformal symmetry does not in itself resolve the question of spurious poles; something more is needed, which we shall reach in due course. The principle we use to achieve this new insight is extremely simple: to represent a conformal symmetry for momentum-space, we use twistor coordinates for momentum-space. 

The following theorem is motivated and guided by dual conformal symmetry: the region coordinates $x_i^a$ correspond to the lines $X_i$ in twistor space appearing in the proof.

{\sc Theorem:} Suppose that for $n\ge 4$, the $2n$ spinors $\pi_{iA'}$, $\pi_{iA}$, satisfy the conditions $$\sum_{1}^{n}\pi_{iA'}\pi_{iA}= 0. $$
We may write $p_{ia} = \pi_{iA'}\pi_{iA}$ so this condition expresses (complex) momentum conservation: $\sum_{i=1}^n p_{ia} = 0$.

Suppose also that the following non-singularity condition holds: that for each $i$, $\pi_{i-1}^{A'}\pi_{iA'} \ne 0\,$ and $\pi_{i-1}^{A}\pi_{iA} \ne 0$. (Equivalently, $S_{i-1,i} = \frac{1}{2}(p_{i-1} + p_i)^2 \ne 0.$)Then there exist $n$ twistors $Z_i^{\alpha}$ and $n$ dual twistors $W_{i\alpha}$ satisfying the following conditions for each $i$:
\begin{trivlist}
\item \quad The $\pi$-part of  $Z_i^{\alpha}$ is $\pi_{iA'}$ and the  $\pi$-part of $W_{i\alpha}$ is  $\pi_{iA},$ \item  \quad $W_{i\alpha}Z_{i-1}^{\alpha} = W_{i\alpha}Z_{i}^{\alpha} = W_{i\alpha}Z_{i+1}^{\alpha} = 0,$ \item  \quad $\epsilon_{\alpha \beta \gamma \delta} Z_i^{\alpha} Z_{i+1}^{\beta} Z_{i+2}^{\gamma} Z_{i+3}^{\delta} \ne 0$. \end{trivlist}
The set of  $\{Z_i^{\alpha}, W_{i\alpha}\}$ satisfying all these conditions is unique up to a linear transformation corresponding to a complex translation in $\mathbb{CM}$.

{\sc Proof:} Let $X_0$ be any line in projective twistor space, corresponding to a finite point $x_0^a$  in  $\mathbb{CM}$. Then specify $Z_0^{\alpha}$, $Z_1^{\beta}$ by the condition that they lie on $X_0$ and have $\pi$-parts $\pi_{0A'}, \pi_{1B'}$ respectively. These points are well-defined and distinct, by the non-singularity condition. Now consider the point $x_0^a + \pi_1^{A} \pi_1^{A'}$, with a corresponding line $X_1$ in twistor space. This line is not coincident with $P_0$, since $\pi_{1A} \pi_{1A'} \ne 0$, and it also represents a finite point in $\mathbb{CM}$.  Note that $Z_1^{\beta}$ lies on $X_1$. Define $Z_2^{\gamma}$ by it lying on $X_1$, and  having $ \pi$-part $\pi_{2C'}$. Clearly $Z_0^{\alpha}, Z_1^{\beta}, Z_2^{\gamma}$ define a plane. Continue in the same way, so that line $X_k$ corresponds to the point  $x_0^a + \sum_{i=1}^{k} \pi_i^{A} \pi_i^{A'}$. Eventually we define a twistor $Z_{n-1}^{\alpha}$ by this means. Now define $X_{n}$ by the same method, and it corresponds to the point $x_0^a + \sum_{i=1}^{n} \pi_{i}^{A} \pi_i^{A'}$. By  the hypothesis on the $2n$ given spinors, this is just  $x_0^a$. Hence $X_{n} = X_0$ and the definition of the $n$ twistors $Z_i^{\alpha}$ is complete and consistent. 

The dual twistors $W_{i\alpha}$ are now defined by the planes containing consecutive triples of the $X_i$. Explicitly, including the scale: 
\begin{equation}W_{i\alpha} = \frac{\epsilon_{\alpha\beta\gamma\delta}Z_{i-1}^{\beta}Z_{i}^{\gamma}Z_{i+1}^{\delta}}  {(I_{\lambda\mu}Z_{i-1}^{\lambda}Z_{i}^{\mu})\, (I_{\nu\sigma}Z_{i}^{\nu}Z_{i+1}^{\sigma})}\, . \label{eqn:formula}
\end{equation}
This is the all-important formula which allows momentum-space amplitudes to be expressed in momentum-twistors. By construction, the $W_{i\alpha}$ satisfy all the orthogonality properties stated. To check that the $\pi$-part of $W_{i\alpha}$ is $ \pi_{iA}$, apply the formula (\ref{eqn:displace}) to the lines $X_{i-1}$ and $X_i$. We obtain for the displacement tensor
\begin{eqnarray}D^{\alpha}_{\beta}&=&\frac{ I^{\alpha\kappa} \epsilon_{\kappa\gamma\delta\theta}Z_{i-1}^{\gamma} Z_{i}^{\delta}Z_{i}^{[\theta} Z_{i+i}^{\phi]} I_{\phi \beta}} {(I_{\lambda\mu}Z_{i-1}^{\lambda}Z_{i}^{\mu})\, (I_{\nu\sigma}Z_{i}^{\nu}Z_{i+1}^{\sigma})} \nonumber \\
&= &I_{\phi \beta}Z_{i}^{\phi} \frac{ I^{\alpha\kappa} \epsilon_{\kappa\gamma\delta\theta} Z_{i-1}^{\gamma} Z_i^{\delta} Z_{i+1}^{\theta}}{{(I_{\lambda\mu}Z_{i-1}^{\lambda}Z_{i}^{\mu})\, (I_{\nu\sigma}Z_{i}^{\nu}Z_{i+1}^{\sigma})}} \nonumber \\ & =&(I_{\phi \beta}Z_{i}^{\phi} )(I^{\alpha\kappa}W_{i\kappa})\, .
\end{eqnarray}
But this must agree with the displacement vector  $\pi_i^{A'} \pi_i^{A}$, so the the $\pi$-part of $W_{i\alpha}$ is $ \pi_{iA}$, as required.

As an alternative to this geometrical proof,  we may state algebraic formulas for the $Z_i^{\alpha}$ and $W_{i\alpha}$ as follows:
\begin{eqnarray}
Z_k^{\alpha} = (i\pi_{kB'} (x_{0}^{AB'} + \sum_{i=1}^{k-1} \pi_{i}^{A}\pi_{i}^{B'}), \,\pi_{kA'}), \nonumber \\
W_{k\alpha} = (\pi_{kA}, - i\pi_{kB}(x_{0}^{BA'} + \sum_{i=1}^{k-1} \pi_{i}^{B}\pi_{i}^{A'}) )\, ,
\end{eqnarray}
and verify that all the conditions are satisfied. This formula brings out how each $Z_i^{\alpha}$ and $W_{i\alpha}$ depends non-linearly on the $2n$ given momentum-spinors.

The line $X_0$ was arbitrary. Choosing a different line is equivalent to a translation in $\mathbb{CM}$, equivalent to a volume-preserving linear transformation of the $Z_{i}^{\alpha}$ which preserves $I_{\alpha\beta}$. We shall find that the $Z_{i}^{\alpha}$ appear  only  through the combinations  $I_{\alpha\beta}Z_{i-1}^{\alpha}Z_{i}^{\beta}$ and $\epsilon_{\alpha\beta\gamma\delta}Z_{i}^{\alpha}Z_{j}^{\beta}Z_{k}^{\gamma}Z_{l}^{\delta}$, which are invariant under just this group of transformations.

For a proof of the uniqueness statement, suppose that $Z_i^{\alpha}$ and $W_{i\alpha}$ are any solutions of these equations. The conditions they satisfy imply the existence of distinct points $x_i^a$, each null-separated from its neighbours, whence the  $Z_i^{\alpha}$ must agree with the  $Z_i^{\alpha}$ as constructed above. The uniqueness fails if the condition  $\epsilon_{\alpha \beta \gamma \delta} Z_i^{\alpha} Z_{i+1}^{\beta} Z_{i+2}^{\gamma} Z_{i+3}^{\delta} \ne 0$ is not imposed. If it is violated, the $x_i^a$ need not be distinct. Let $X_0$ be any line corresponding to a finite point in $\mathbb{CM}$;  there exist points $Z_i^{\alpha}$ {\em all}  on $X_0$, and $W_{i\alpha}$ planes {\em all} containing  $X_0$, satisfying all the other conditions.

The amplitudes are generally singular if any $S_{i-1, i} $ vanishes, so this exclusion does not impede our ability to map momentum-space amplitudes into the new momentum-twistor coordinates. In the reverse direction, we shall find that the non-vanishing of $\epsilon_{\alpha \beta \gamma \delta} Z_i^{\alpha} Z_{i+1}^{\beta} Z_{i+2}^{\gamma} Z_{i+3}^{\delta}$ is also a natural condition on the amplitudes. For the analysis of infra-red divergences and soft limits, however, it may be profitable to scrutinise in finer detail the way that the correspondence between momenta and the  $Z_i^{\alpha}$ fails to hold in this singular region. 

Having excluded this singular region, the $W_{i\alpha}$ are completely defined by the  $Z_i^{\alpha}$, which are themselves defined, up to the freedom stated, by the $2n$ spinors. So  we have a good encoding of the given $2n$ spinors in terms of $n$ twistors --- a coding which {\em absorbs} the momentum-conservation condition. The $\pi$-parts of dual twistors are replaced by $\omega$-parts of twistors (or, equally well, the other way round). The \hbox{4-vector} condition implied by the sum $\sum_{1}^{n}\pi_i^{A'}\pi_{iA}= 0$ has been converted into the 4-vector freedom in the choice of the $Z_i^{\alpha}$. This is just the same freedom as in the `region space' of $x^a$ on which dual conformal symmetry is defined, with the slight difference that in the twistor geometry there is no reason to consider the momenta as being real. 

It is convenient to introduce some notation. We have already  used the convention of angle-brackets and square brackets for spinor products, thus
$$ \langle 12 \rangle = \pi_1^{A'} \pi_{2A'} = I_{\alpha \beta}Z_1^{\alpha}Z_2^{ \beta}\, ,$$
$$[12] = \pi_1^{A} \pi_{2A} = I^{\alpha \beta}W_{1\alpha}W_{2 \beta}\, .$$
Now define also the conformally invariant (actually SL4C invariant) objects
\begin{equation}
\langle 1234 \rangle = \epsilon_{\alpha \beta \gamma \delta}Z_1^{\alpha}Z_2^{ \beta}Z_3^{\gamma}Z_4^{ \delta}\, ,\end{equation}
\begin{equation}[1234] = \epsilon^{\alpha \beta \gamma \delta}W_{1\alpha}W_{2 \beta}W_{3 \gamma}W_{4 \delta}\, .\end{equation}
Then from applying formula (\ref{eqn:formula}),
\begin{equation}[12] = \frac{\langle 0123\rangle } {\langle 01\rangle \langle 12 \rangle \langle 23 \rangle}\, ,
\end{equation}
\begin{equation}s_{12} = \frac{1}{2}(p_1 + p_2)^2 =  \langle 12\rangle [12] = \frac{1}{2}(x_0 - x_2)^2  = \frac{\langle 0123\rangle } {\langle 01\rangle  \langle 23 \rangle} \, .
\end{equation}
The property of the amplitudes, essential to dual conformal invariance, is that the momenta only enter through sums of {\em consecutive} momenta (consecutive, that is, in the ordering defined by the colour-trace). In the momentum-twistor representation, this is reflected in the simplicity of the further invariants
\begin{equation}
S_{123} = \frac{1}{2} (x_0 - x_3)^2 =   \frac{\langle 0134\rangle } {\langle 01\rangle  \langle 34 \rangle} \, , \, S_{1234} = \frac{1}{2} (x_0 - x_4)^2 =   \frac{\langle 0145\rangle } {\langle 01\rangle  \langle 45 \rangle} \, ,
\end{equation}
and so on. 

\subsection{Translating the amplitude}

Now we translate the momentum-space expression for  $A(1^- 2^- 3^- 4^+ 5^+ 6^+)$ into the new momentum-twistor coordinates. Formula (\ref{eqn:formula}) shows that:
\begin{equation} [4|5+6|1\rangle = \frac{\langle 1345\rangle}{\langle 34\rangle \langle 45 \rangle } ,\,\,\, [6|1+2|3\rangle = \frac{\langle 1356\rangle}{\langle 56\rangle \langle 61 \rangle } , \,\,\,[2|3+4|5\rangle = \frac{\langle 1235\rangle}{\langle 12\rangle \langle 23 \rangle } \, .
\end{equation} 
It follows that the two terms of (\ref{eqn:first}) become:
\begin{equation}\frac{\langle 12\rangle^4 \langle 23\rangle^4}{\langle 12\rangle \langle 23\rangle  \langle 34\rangle  \langle 45\rangle  \langle 56\rangle  \langle 61\rangle} \frac{1}{\langle 1235\rangle} \left(\frac{\langle 1345\rangle^{3}}{  \langle 2345\rangle \langle 1234\rangle \langle 1245\rangle} - \frac{\langle 1365\rangle^{3}}{\langle 2365\rangle \langle 1236\rangle \langle 1265\rangle}\right)\, .
\label{eqn:spurious}
\end{equation} 
The vital feature of this expression is the spurious pole $\langle 1235\rangle$. The simple anti-symmetry of this transformed expression makes it much easier to see why it is indeed a removable singularity. To do this, note the  identity   which holds for all $a, b, c, d, e, f$:
\begin{equation}\langle abcd\rangle \langle abef\rangle + \langle abce\rangle \langle abfd\rangle + \langle abcf\rangle \langle abde\rangle = 0\, .
\end{equation}
(This is simply the spinor identity, applied to the line defined by $Z_a^{[\alpha}Z_b^{\beta]}$.)
Applying it with $a=3, b=5, c=1, d=2, e=4, f=6$, we have
\begin{equation}\frac{\langle 1345\rangle}{\langle 2345\rangle} = \frac{\langle 1365\rangle}{\langle 2365\rangle} +   \frac{\langle 1235\rangle \langle 4635\rangle}{\langle 2345\rangle \langle 2365\rangle }
\end{equation}
so that when $\langle 1235\rangle =0$, $\langle 1345\rangle/\langle 2345\rangle = \langle 1365\rangle/\langle 2365\rangle$. Likewise  $\langle 1345\rangle/\langle 1234\rangle = \langle 1365\rangle/\langle 1236\rangle$ and $\langle 1345\rangle/\langle 1245\rangle = \langle 1365\rangle/\langle 1265\rangle$.
Thus when $\langle 1235\rangle =0$, the bracketed expression in (\ref{eqn:spurious}) vanishes, so  $\langle 1235\rangle$ is a removable singularity.

One could use this algebra to give an explicit formula for the complete amplitude, in which the spurious pole is absent. Such formulas, and more generally the use of the new coordinates, might be useful for practical calculation. (They do not make possible anything that could not, in principle, have been expressed in the original spinors. But the absorption of the $\delta$-function constraints into the coordinates makes for much simpler algebra.) However, this is not the main point of this note. Our goal is to show a simple {\em geometrical} characterisation of spurious poles. Equivalently, we shall show why the  terms can be summed into a single object, quite unlike the listing of terms yielded by the BCFW recursion relation.

\section{New integrals in momentum-twistor space}

A first crucial observation is that the conformal invariant
$$\frac{\langle 1345\rangle^{3}}{\langle 1235\rangle  \langle 2345\rangle \langle 1234\rangle \langle 1245\rangle}$$
can be expressed as the result of the projective dual twistor integral
\begin{equation}\int_{T_{1345}} \frac {6 }{(W.Z_2)^4} \,\, DW
\label{eqn:cp3integral}
\end{equation}
where $T_{1345}$ defines a 3-dimensional contour with boundaries on $W.Z_1=0, W.Z_3=0, W.Z_4=0, W.Z_5=0$.

 As many-dimensional complex integrals with boundary are not universally familiar, it may be helpful to sketch the main features of (\ref{eqn:cp3integral}), starting with a one-dimensional analogue, viz.\  a line integral in $\mathbb{CP}^1$ . We consider
\begin{equation}\int_{L_{12}} \frac {1 }{(\pi . \alpha)^2} \,\, D\pi
\end{equation}
where $L_{12}$ denotes a path in $\mathbb{CP}^1$ with end-points at $\pi = \sigma_1, \pi=\sigma_2$ (or equivalently, at the points where $\pi.\sigma_1 =0, \pi.\sigma_2 =0$). One way of performing the integral is to map the  $\mathbb{CP}^1$, minus the point $\pi = \alpha$, into $\mathbb{C}$, by  $z=  \pi.\lambda/ \pi.\alpha$ for some $\lambda \ne \alpha$.  The point $\alpha$ can be thought of as being sent to infinity on the Riemann sphere. The integral becomes 
$$\int_{s_1}^{s_2} dz$$
where ${s_i} = ( \sigma_i.\lambda)/(\alpha.\sigma_i)(\alpha.\lambda)$, thus yielding 
$( \sigma_1.\sigma_2)/(\alpha.\sigma_1)(\alpha.\sigma_2)$. 

The path in the $z$-plane may take any form whatever, but there is a particular representative contour given by $\{z= uz_1 + (1-u)z_2\, | \, u\in [0,1]\}$. We could use the same `real line segment' definition in the original projective integral by choosing the path $L_{12}$ to be $\{\pi= u\sigma_1 + (1-u)\sigma_2\, | \, u\in [0,1]\}$. 

Note that the spinor identity can be written as $$(\alpha.\sigma_3)( \sigma_1.\sigma_2) + (\alpha.\sigma_1)( \sigma_2.\sigma_3) + (\alpha.\sigma_2)( \sigma_3.\sigma_1) = 0\, .$$
Hence
\begin{eqnarray}
&&\int_{L_{12} + L_{23} + L_{31} } \frac {1 }{(\pi . \alpha)^2} \,\, D\pi  \nonumber \\
&=&\int_{L_{12}} \frac {1 }{(\pi . \alpha)^2} \,\, D\pi + \int_{L_{23}} \frac {1 }{(\pi . \alpha)^2} \,\, D\pi + \int_{L_{31}} \frac {1 }{(\pi . \alpha)^2} \,\, D\pi \nonumber \\ 
&=&
( \sigma_1.\sigma_2)/(\alpha.\sigma_1)(\alpha.\sigma_2) + ( \sigma_2.\sigma_3)/(\alpha.\sigma_2)(\alpha.\sigma_3) + ( \sigma_3.\sigma_1)/(\alpha.\sigma_3)(\alpha.\sigma_1)\nonumber \\  &=& 0
\label{eqn:sigmas}
\end{eqnarray}
showing that the spinor identity is equivalent to the fact that the triangular path $L_{12} + L_{23} + L_{31}$ is closed, which we can write as $L_{12} + L_{23} + L_{31}= 0$. This also gives another way of expressing the simplest example of cancelling spurious poles, since ({\ref{eqn:sigmas}) is nothing but  the partial fraction decomposition ({\ref{eqn:zvw}) expressed in $\mathbb{CP}^1$ coordinates.

Now we make the analogous definition of a {\em tetrahedral} contour for (\ref{eqn:cp3integral})
by solving for the vertices $V_{134}, V_{135} V_{145}, V_{345}$, i.e.\ the points in projective $W$ space which satisfy three boundary conditions, and then defining the contour:
$$\{W = x V_{134} + y  V_{135} + z V_{145} + (1-x-y-z) V_{345} \, | \, x \in [0,1], y \in [0,1], z \in [0,1]\}\,.$$ 
It may easily be verified that this yields the stated result.
As in the one-dimensional analogue, this is only a representative contour. But the properties of homomorphic functions ensure that the result of the integral is independent of the representative. Although the ambient space is 6-real-dimensional, it is not misleading to consider the integral as related to the {\em volume} of a tetrahedron. Because the incidence properties of planes, lines and points in complex projective space are just the same as in real spaces, it is also legitimate to picture the contour  integrals using figures in three real dimensions, and we shall do so in the following.

We now note that likewise \begin{equation}\frac{\langle 1365\rangle^{3}}{\langle 1235\rangle  \langle 2365\rangle \langle 1236\rangle \langle 1265\rangle} = \int_{T_{1365}} \frac {6 }{(W.Z_2)^4} \,\, DW
\label{eqn:cp3integral6}
\end{equation}
where $T_{1365}$ has bounding faces $W.Z_1=0, W.Z_3=0, W.Z_5=0, W.Z_6=0.$

In both integrals, the necessary condition $\langle 1235\rangle \ne 0$ can be interpreted as the condition that the vertex $V_{135}$ is a finite point when the bounding face corresponding to $Z_2^{\alpha}$  is sent to infinity.

\subsection{Why spurious poles cancel: spurious boundaries}

We have neglected questions of sign in the preceding discussion, (and the {\em overall} sign will continue to neglected)  but more precisely, we have  tetrahedral contours  equipped with an {\em orientation.} We shall use the sign of the permutation to indicate relative orientation, thus writing $T_{1365} = - T_{1356}.$ Then the difference between (\ref{eqn:cp3integral}) and (\ref{eqn:cp3integral6}) is equivalent to integrating over $ T_{1345} - T_{1365} = T_{1345} + T_{1356}$. This is a new polyhedron $P_6 = T_{13[46]5}$  with 6 vertices, 9 edges and 5 faces (Figure 2). The  vertex $V_{135}$ is absent. It follows that the combined integral, giving the amplitude, remains finite even when  the vertex $V_{135}$ is at infinity. This explains geometrically why the pole $\langle 1235\rangle$ no longer appears in the amplitude.

\begin{figure}[h]
\begin{center}
\includegraphics[width=186.px]{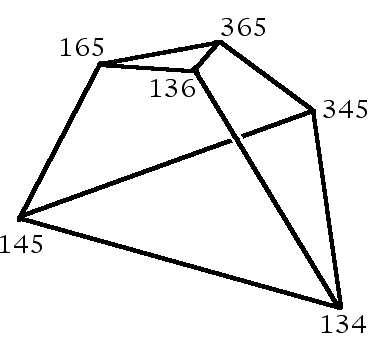}
\caption{ The polyhedron $P_6 = T_{13[46]5}$, with 6 vertices, 9 edges and 5 faces.}
\end{center}
\end{figure}

Our guiding idea is that {\em spurious poles arise from spurious boundaries}.  If we consider the amplitude to be given by the integration of $(W.Z_2)^{-4}$ over $P_6$, which has no boundary at  the vertex $(1235)$, the spurious pole never arises. The spurious poles only arise from the {\em representation} of $P_6$ as the difference of $T_{1345}$ and $T_{1365}$, which requires the insertion of a spurious boundary.

An observation as elementary as school geometry now allows a marvellous application of this identification of spurious boundaries. Note that  $P_6$ can be decomposed in a quite different way into the sum of {\em three} tetrahedra:
$$P_6 = T_{1346} + T_{3546} + T_{5146}$$
This is most easily seen in the dual picture (Figure 3), where these three tetrahedra meet on their common edge $\{46\}$.

\begin{figure}[h]
\begin{center}
\includegraphics[width=122.px]{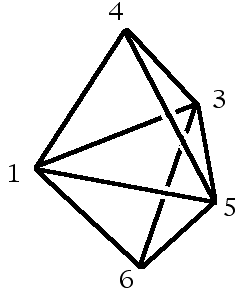}
\caption{ In the dual representation,  $P_6$ appears as the {\em join} of two tetrahedra, with 5 vertices, 9 edges and 6 faces. }
\end{center}
\end{figure}

It follows that the volume of this polytope can also be written as
\begin{equation}
\frac{1}{\langle 1246\rangle \langle 2346\rangle} \left(\frac{\langle 1346\rangle^{3}}{  \langle 1234\rangle \langle 1236\rangle}\right) + \frac{1}{\langle 2346\rangle \langle 2546\rangle} \left( \frac{\langle 3546\rangle^{3}}{ \langle 2345\rangle \langle 2356\rangle} \right) $$ $$+ \frac{1}{\langle 1246\rangle \langle 2546\rangle} \left(\frac{\langle 5146\rangle^{3}}{\langle 1245\rangle \langle 1256\rangle}\right)
\label{eqn:threeterm}
\end{equation}
But this expression corresponds exactly to the formula (\ref{eqn:second}) for the amplitude. Thus twistor geometry reduces the hexagon identity almost to triviality in the special case of split-helicity.

In this case it is the vertices $V_{146}, V_{346}, V_{546}$ that correspond to spurious poles. (Again, this is more easily seen in the dual picture, where these vertices correspond to the internal faces which split the polyhedron into three parts.) 

The splitting of the polyhedron in two different ways can be stated in a more advanced and elegant geometrical form. The five-term identity stating the equivalence of (\ref{eqn:first}) and (\ref{eqn:second}) corresponds to the fact that the five tetrahedral hyperfaces of a 4-dimensional simplex form a closed boundary. It is a higher-dimensional analogue of the closed triangle in one dimension, $L_{12} + L_{23} + L_{31} = 0$. Algebraically, the spinor identity is equivalent to   $T_{[a}\epsilon_{bc]} = 0$ for any $T$ in two dimensions, and the five-term identity to $T_{[a}\epsilon_{bcde]} = 0$ for any $T$ in four dimensions.

It is natural to refer to `spurious' and `physical' vertices, and we shall do so in what follows. The definition of a physical vertex is that it is given by a quadruple which can be put in the form $(j, j+1, k, k+1)$ for some $(j,k),$ and so corresponds to some physical pole. All the others are spurious. We have shown that $P_6$ has only physical vertices, but we want to ensure that this is not some special effect which applies to  $n=6$ alone.

 \subsection{Generalisation to more than six fields}

In fact this geometric description readily generalises to split-helicity $n$-field amplitudes in the NMHV sector.  

As preparation for doing this, and to complete the picture, we should first go back to the case of NMHV for {\em five} fields in the split-helicity case. It is easy to verify that  $A(1^- 2^- 3^- 4^+ 5^+)$ corresponds to integrating over the polyhedron $P_5 = T_{1345}$. That is,
$$A(1^- 2^- 3^- 4^+ 5^+) = \frac{[45]^4}{[12][23][34][45][51]} = \frac{\langle 12\rangle^4 \langle 23\rangle^4}{\langle 12\rangle \langle 23\rangle \langle 34\rangle \langle 45\rangle \langle 51\rangle}
\int_{P_5} (W.Z_2)^{-4}  DW \, .$$
Note that the change in going from five to six fields is very simple:  
(A) the denominator factor expands from five to six terms and (B)  the  new polyhedron is the sum of the polyhedron for $A(1^- 2^- 3^- 4^+ 5^+)$ and the polyhedron for $A(1^- 2^- 3^- 5^+ 6^+)$. The same simple jump will occur in going up to 7 and more fields, and it is worth exploiting its simplicity to avoid writing down long expressions analogous to (\ref{eqn:first}) and (\ref{eqn:second}). One way of doing this is through the twistor diagram representation of the amplitudes. Nothing in what follows depends on using the twistor diagram representation, but it is helpful in showing graphically the combinatorial aspects of the amplitudes. Furthermore, the feature of twistor diagrams that we shall use now is one that Arkani-Hamed et al.\ (2009) have pointed out as being of special interest.

The two-term formula (\ref{eqn:first}) may be represented by the diagrams in Figure 4.

\begin{figure}[htbp] %  figure placement: here, top, bottom, or page
   \centering
   \includegraphics[width=414px]{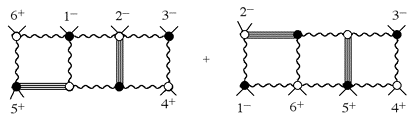} 
   \caption{Twistor diagrams for the two-term representation of the split-helicity amplitude $A(1^- 2^- 3^- 4^+ 5^+ 6^+)$}
   \label{fig:ex4}
\end{figure}

Here we are using the definition of diagrams as given in (Hodges 2005a). The wavy lines are boundaries and the quadruple lines are quadruple poles. In the super-symmetric extension developed in (Hodges 2005b, 2006), these become super-components of diagrams in which all the edges represent super-boundaries. However, the restriction to split-helicity makes it appropriate to revert to the more elementary form. (The same features will be found in the diagrams defined for the analogous split-signature amplitudes by Arkani-Hamed et al.\ (2009), although their wavy lines have a different meaning!)

Note that the first  of these diagrams can be seen as the result of using boundary-lines to attach the 6-field  to the  twistor diagram for $A(1^- 2^- 3^- 4^+ 5^+)$  and the second as the result of attaching the 4-field to $A(1^- 2^- 3^- 5^+ 6^+)$  similarly. This is no coincidence, and it illustrates a general simplifying rule. This attachment can be seen as a special case of the BCFW joining process, namely that which arises when one of the sub-amplitudes is a 3-amplitude. There are two cases of this: joining an extra field to two twistor vertices (black), or to two dual twistor vertices (white). Arkani-Hamed et al.\ (2009) call this adding a black or white triangle respectively, and give an interpretation of the attachment process as the inverse of a soft limit. We may verify that when it is a {\em black} triangle that is added, the effect on the amplitude is extremely simple in the new coordinates: the denominator factors are expanded appropriately, and the polyhedron of integration remains {\em unchanged.} This may seem surprising, because the `momentum shift' seems to have been neglected. The explanation is that it has gone into the re-definition of the momentum-twistors in the new context. The polyhedron remains the same, but its interpretation in terms of the external momenta changes in exactly the right way.

That $P_6 = T_{1345} + T_{1356}$ follows immediately from this observation.
This method takes advantage of the fact, emphasised by Arkani-Hamed, that twistor diagrams make manifest the many different ways in which a term can arise by the composition of sub-amplitudes.  
 
The twistor diagrams for the 3-term expansion (\ref{eqn:second}) can also be related to the 5-field amplitude, but not quite so simply: it needs the supersymmetric extension of the `attachment' principle.

\subsection{NMHV split-helicity amplitudes for seven fields}

In this light we consider $A(1^- 2^- 3^- 4^+ 5^+ 6^+7^+)$. The BCFW expansion, pivoting on (7,1), leads to only two non-vanishing terms:
\begin{eqnarray}  A( k^+1^-2^-3^- 4^+5^+)\circ A(6^+ 7^+ k^- ) \nonumber \\
+  A( k^+1^- 2^- )\circ A(3^- 4^+ 5^+6^+7^+k^-)\, .
 \end{eqnarray}
Use of  the twistor diagrams streamlines the identification of the corresponding polyhedra. The first term is the so-called `homogeneous' term, and corresponds to attaching the 7-field to the NMHV diagram for the other six fields, and so to integrating over the polyhedron $P_6$. The second term corresponds  to attaching fields 4 and 5 to the five-field diagram for $A(1^-2^-3^-6^+7^+)$, and so to integrating over  $T_{1367}$. Thus $$A(1^- 2^- 3^- 4^+ 5^+6^+7^+)  = \frac{\langle 12\rangle^4 \langle 23\rangle^4}{\langle 12\rangle \langle 23\rangle \langle 34\rangle \langle 45\rangle\langle 56\rangle\langle 67\rangle \langle 71\rangle}
\int_{P_7} (W.Z_2)^{-4}  DW $$

where the  polyhedron $P_7$ is produced by adding a further tetrahedron to $P_6$: 
\begin{equation}P_7 = P_6 + T_{1367} = T_{1345} + T_{1356} + T_{1367}\, .
\label{eqn:p7}
\end{equation}
$P_7$ has 6 faces, 12 edges and 8 vertices, but it is not like a cube, for two faces are pentagonal and two triangular.  For a picture of it, take $P_{6}$ and truncate the vertex $V_{136}$ (which is now spurious) with the 7-plane. The result is a degenerate pentagonal prism: two pentagons, joined on one edge, and with the remaining vertices of the pentagons also joined by edges.

In the dual picture we have 8 faces, 12 edges and 6 vertices, but not an octahedron. It is like an octahedron with a quadrant removed. Take an octahedron with vertices 1 and 3 opposite, 4 and 6 opposite, and 5 and 7 opposite. Now cut away the tetrahedron $T_{1374}$. Explicitly, note that  $P_7 = T_{46[57][13]} - T_{1374}$, 
where  $T_{13[46][57]} = T_{1345} + T_{1356} + T_{1367} + T_{1374} = T_{46[13][57]} = T_{57[13]46]}$ has octahedral symmetry. These statements of alternative representations indicate the rich variety of ways in which the amplitude can be re-expressed. The identities can also all be seen as expressions of the hexagon identity, i.e.\ to the 3-dimensional boundary of a 4-dimensional simplex being closed.

 All the vertices in $P_7$ are of form $(1, i, i+1)$ or $(3, i, i+1),$ and so correspond to physical poles of the particular form $\langle 12,i,i+1\rangle$ or $\langle 23,i,i+1\rangle$. The other possible physical poles do not occur, because of the very special nature of the  split-helicity case.

The different expressions which arise from using other pivots can now be interpreted simply as different representations of the same polyhedron. Using (56) as pivots, for instance, gives rise to a BCFW expansion with three non-vanishing terms:
\begin{eqnarray} A(1^- 2^- 3^- 4^+ 5^+k^+)\circ A(k^- 6^+7^+)\nonumber \\
 + A(2^- 3^- 4^+ 5^+k^+)\circ A(k^- 6^+7^+1^- ) \nonumber \\
 +  A( 3^- 4^+ 5^+k^-)\circ A(k^+ 6^+7^+1^-2^- ) \, .
 \end{eqnarray}
The first (`homogeneous') term is equivalent to attaching the 6-field to \newline$ A(1^- 2^- 3^- 4^+ 5^+ 7^+).$ The second and third terms correspond to $T_{5167}$ and $T_{3567}$. These can readily be seen in the twistor diagram representation as the effect of attaching the 4-field to two six-field terms, each already known from the twistor diagrams corresponding to (\ref{eqn:second}).

It is easy to verify that indeed $P_7 = T_{13[47]5} +  T_{5167} + T_{3567} $.  

\subsection{NMHV split-helicity amplitudes for $n$ fields}

We can generalise these observations as follows. Let $n \ge 5$. Then 
\begin{equation}
A(1^- 2^- 3^- 4^+ 5^+ 6^+\ldots n^+)  = \frac{\langle 12\rangle^4 \langle 23\rangle^4}{\langle 12\rangle \langle 23\rangle \langle 34\rangle\ldots \langle n1\rangle}
\int_{P_n} (W.Z_2)^{-4}  DW
\end{equation}
where the polyhedron $P_n$ is defined by
\begin{equation}P_n = P_{n-1} + T_{1,3,n-1, n} = \sum_{i=4}^{n-1} T_{1,3,i, i+1} \, .
\end{equation}
The polyhedron $P_n$ has $n-1$ faces, $3n-9$ edges and $2n-6$ vertices. It can be pictured by a continuation of the process of truncation that gave the polyhedron for $n=7$. Truncate the vertex $V_{137}$ with the 8-plane, then  $V_{138}$ with the 9-plane, and so on, at each stage removing the vertex that becomes spurious in the larger context. This result is again a degenerate prism: a polyhedron given by two $(n-2)$-gons, joined on one edge, and the remaining vertices joined, in order, by edges.

\begin{figure}[htbp] %  figure placement: here, top, bottom, or page
   \centering
   \includegraphics[width=143px]{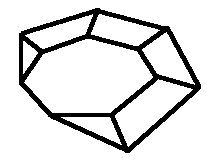} 
   \caption{Polyhedron $P_9$}
   \label{fig:ex5}
\end{figure}

For the dual picture, take a $(n-3)$-gon with vertices at $4, 5, 6 \ldots n$. Take vertices 1 and 3, and join each of these to all the vertices of the $(n-3)$-gon. Now remove from the resulting polyhedron  the tetrahedral slice $T_{134n}.$

The proof of this correspondence rests on performing the BCFW recursion with pivot (3,4). As in the case of 7 fields, we find that for $n$ fields,  there is only one term to add on to the `homogeneous' term which corresponds to the polyhedron already established for $n-1$. Moreover this additional term can always be identified with a tetrahedron by the simple `attachment' rule.

These specific results shed some new light on the short `split-helicity' formulas which have attracted attention in the past. In general, with a splitting into $r$ and $(n-r)$ consecutive helicities of the same type, there exist short formulas of length $\frac{(n-4)!}{(r-2)!(n-r-2)!}$ terms (Hodges 2005b).  We have studied the case $r=3$. Study of the cases $r > 3$ may assist in developing the theory beyond NMHV, and this is one possible direction of generalisation.

\section{Supersymmetric extension to all helicities}

But it is also desirable that this geometric formalism should be developed to encompass all helicity configurations, and not just split-helicity. It is clear that we should look for a  super-symmetric extension. The principle, proceeding  by analogy with the super-symmetric extension of twistor diagrams (Hodges 2005b), is that of replacing all the boundaries by super-boundaries, and  the $(W.Z_2)$ quadruple pole also by a super-boundary. This restores the full symmetry in a larger (super)-space.  When restricted to the split-helicity case, this super-boundary then acts as if it were a quadruple pole, thus reducing the remaining integration to one over the polyhedron $P_n$.

The definition of the supersymmetric extension of the momentum-twistors is beyond the scope of this introductory note, but a brief sketch will be given of what it means to replace a boundary by a super-boundary. The mechanism is the same as in  (Hodges 2005b), which uses super-boundary contours. It uses some simple geometry which goes back to the inhomogeneous twistor integrals introduced in (Hodges 1985).

First note the immediate generalisation of (\ref{eqn:cp3integral}) from $\mathbb{CP}^3$ to $\mathbb{CP}^4$:
\begin{equation}\int_{H_{12345}} \frac {24}{(w.z_0)^5} \,\, Dw = \frac{\langle 12345\rangle^{4}}{\langle 01234\rangle  \langle 01235\rangle \langle 01245\rangle \langle 01345\rangle \langle 02345\rangle} 
\label{eqn:cp4integral}
\end{equation}
where $w, z_{i}$ are $\mathbb{CP}^4$ variables, $Dw$ the natural projective 4-form, ${H_{12345}}$ is the simplex with five hyperfaces on $w.z_1=0, \ldots w.z_5=0$, and $\langle 12345\rangle$ stands for $\epsilon_{\alpha \beta \gamma \delta \theta}z_1^{\alpha}z_2^{\beta}z_3^{\gamma}z_4^{\delta}z_5^{\theta}$.

If  $\mathbb{CP}^4$, minus the hyperplane $\{ w | (w.z_0) = 0\}$, is mapped onto dual twistor space $\mathbb{T^{*}} \simeq \mathbb{C}^4$,
the  $\mathbb{CP}^4$ integral (\ref{eqn:cp4integral}) becomes:
$$24 \int_{H} D^4 W$$
where $H$ is a 4-dimensional contour with boundaries on the hyperfaces of a simplex, which may be given in the form
$$W.Z_1 + c_1=0, W.Z_2 + c_2=0, W.Z_3 + c_3=0, W.Z_4 + c_4=0, W.Z_5 + c_5=0\,.$$
The result, which can now be interpreted as \ (24 times the) complex 4-volume of this simplex, is
\begin{equation}\frac { \left| \begin{array}{ccccc} c_{1} & Z_1^0 & Z_1^1 & Z_1^2 & Z_1^3\\ c_{2} & Z_2^0 & Z_2^1 & Z_2^2 & Z_2^3\\ c_{3} &  Z_3^0 & Z_3^1 & Z_3^2 & Z_3^3\\ c_{4} &  Z_4^0 & Z_4^1 & Z_4^2 & Z_4^3\\ c_{5} &  Z_5^0 & Z_5^1 & Z_5^2 & Z_5^3 \end{array} \right| ^{4}}
{\langle 1234\rangle  \langle 1235\rangle \langle 1245\rangle \langle 1345\rangle \langle 2345\rangle} \, .
\label{eqn:determinants}
\end{equation}
This calculation could have be done directly in  $\mathbb{C}^4$ without reference to $\mathbb{CP}^4$, for this formula is simply the complexification of some elementary Euclidean geometry. But the projective description may be helpful by suggesting an analogy with the 3-dimensional tetrahedral contour described in (\ref{eqn:cp3integral}) where a plane in projective twistor space is sent to infinity in  $\mathbb{C}^3$ coordinates.

The formula (\ref{eqn:determinants})  is simply a rational function of all the variables, and in particular a quartic polynomial in the $c_i$. If we have linear relationships between the 4-volumes of such simplexes (such as will be discussed in the next section), then they define identities between the corresponding rational functions. Multiplying up by the denominators, they are simply identities between polynomials. 

Super-boundaries are formally defined as boundaries on $W.Z_i + \phi.\psi_i=0$, where $\phi$ is the anticommuting part of the dual supertwistor $W$, and $\psi_i$ is the anticommuting part of the supertwistor $Z_i$. The result  of the integration over a simplex in super-space, with hyperfaces given by five such super-boundaries, can thus be unambiguously defined by replacing the $c_i$ in (\ref{eqn:determinants}) by  $(\phi.\psi_i)$. Polynomial identities will be preserved in this replacement, so that linear relationships between 4-polytopes will imply the analogous relationships between super-volumes in super-space. 

After making this replacement in (\ref{eqn:determinants}), the formal integration over $\phi_1,\phi_2,\phi_3,\phi_4 $ is immediate and leads to 
\begin{equation}
\frac {(\sum_{i=1}^{5} \psi_i \langle i+1, i+2, i+3, i+4\rangle) ^{4}}
{\langle 1234\rangle  \langle 1235\rangle \langle 1245\rangle \langle 1345\rangle \langle 2345\rangle} \, .
\end{equation}

as (24 times) the super-volume of the super-simplex. This gives the geometric basis for  the supersymmetric theory. Informally, the construction is like treating the `fermionic' part of a supertwistor as a single extra dimension, choosing coordinates in which `classical' twistors, those  with no fermionic part, are sent to infinity, and then measuring a super-volume in the resulting space.
 
\subsection{Polytopes with dihedral symmetry}

The results of Drummond et al.\ (2008) virtually guarantee the validity of such a super-symmetric extension of the dual conformal symmetry, for the individual terms of the BCFW recursion expansion. We now turn to another feature of four-dimensional geometry, namely that 4-polytopes $H_n$ exist with exactly the right properties to express the {\em summation} of those terms into single expressions for the NMHV super-amplitudes, eliminating spurious poles. We shall show:
\begin{trivlist}
\item  (1) The polytopes $H_n$ have dihedral symmetry. 
\item (2) Their vertices correspond precisely to the `physical' vertices, with all spurious vertices eliminated.
 \item(3) The many representations of NMHV super-amplitudes correspond to the many representations of the 4-polytopes as sums of simplexes. 
 \item(4) Each of their $n$ hyperfaces is a copy of the polyhedron $P_n$.
\end{trivlist}

First define $H_5(12345)$ in $\mathbb{CP}^4$ as the (oriented) 5-simplex with hyperfaces (1,2,3,4,5), in analogy with the three-dimensional tetrahedra.
Then we can define: 
\begin{equation}H_6(123456) = H_5(12345) + H_5(12356) + H_5(61345)\, . \label{eqn:h6}
\end{equation}
$H_6(123456)$ has dihedral symmetry, as may be seen simply by listing
its  6 hyperfaces $ (1, 2, 3, 4, 5, 6)$, 15 faces (all 15 possible pairs), 18 edges (all 20 possible except (135) and (246)), and the 9 `physical' vertices:$$(1234, 2345, 3456,  4561, 5612, 6123, 1245, 2356, 3461).$$
Note that Euler's formula, $H-F+E-V=0,$ is satisfied. 

The six-term identity $H_6(123456) =  H_6(234561)$, which expresses the dihedral symmetry, may be expressed as saying that the four-dimensional boundary of a simplex in five dimensions is closed.

Each hyperface is a polyhedron, whose structure can be seen from  the vertices, lines and faces within it. For example, the 2-hyperface has six vertices  $$(1234, 2345, 5612, 6123, 1245, 2356)$$ 
and a structure of edges and faces  which is just that of the  polyhedron $P_6$.

We next define a polytope with 7 hyperfaces:
\begin{equation}H_{7}(1234567) = H_6(123456) +H_{5}(67123) + H_5(67134)+ H_5(67145)\, .
\end{equation} 
Enumeration shows that that the polytope has dihedral symmetry. The first term $H_{6}(123456)$ gives a polytope with 6 vertices corresponding to physical singularities: (1234), (1245), (2345), (3456), (1256), (2356), but 3 which are spurious: (1236), (1346), (1456). These spurious vertices are cancelled by vertices of $H_{5}(67123) , H_{5}(67134), H_{5}(67145)$ respectively. Likewise one may verify that $H_{5}(67123) , H_{5}(67134), H_5(67145)$ supply the further 8 `physical' vertices, whilst 2 further spurious vertices (6714), (6713) cancel between themselves.
The resulting polytope has 14 vertices, 28 edges, 21 faces and 7 hyperfaces, and despite the asymmetry of its definition, has complete dihedral symmetry.  Each hyperface is like the `degenerate prism' polyhedron $P_7$. 

It is straightforward to give an inductive rule for the sum of $\frac{1}{2}(n-4)(n-3)$ simplexes, which generalises the case of $n=7$:
\begin{equation}
 H_{n}(123...n) = H_{n-1}(123\ldots n-1) +  \sum_{j=1}^{n-4}H_{5}(n-1, n, 1, j+1, j+2) \, .
\label{eqn:hn} \end{equation}
The leading term generates $(n-4)$ spurious vertices, namely
$$(n-3, n-2, n-1, 1), (n-4, n-3, n-1, 1), \ldots (3, 4, n-1, 1), (2, 3, n-1, 1).$$
The $j$th of these is cancelled by a spurious vertex from the $j$th of the tail terms. 

The $j=1$ tail term has one further spurious vertex, namely $(n-1, n, 1, n-3)$, which cancels with the same vertex arising in the $j=2$ term. The $j=n-4$ term also has just one further spurious vertex, $(n-1, n, 1, 3)$, which cancels with the vertex in the $j=n-5$ term. Otherwise, for each $j=3,4,5\dots n-5$, the $j$th term have three spurious vertices, one cancelling with the leading term, and the others cancelling with vertices from the $(j-1)$th and $(j+1)$th terms. 

The polytope $H_n$ has $\frac{1}{2}n(n-3)$ vertices (corresponding to the physical poles), $n(n-3)$ edges, $\frac{1}{2}n(n-1)$ faces and $n$ hyperfaces. Each hyperface is a polyhedron $P_n$, as may readily verified from the recursive definition of $H_n$. At each stage, its intersection with the 2-hyperface agrees with the recursive definition of $P_n$. 

\subsection{The combinatorics of BCFW}

We can immediately make a connection with the combinatorics of the BCFW calculation. Taking the pivots as (71), the general expansion of the 7-field amplitude is of form
$$ A(67k)\circ A(k12345) + A(567k)\circ A(k1234) + A(4567k)\circ  A(k123) + A(34567k)\circ A(k12)\,.$$
The twistor diagram representation of the expansion makes the combinatorial form of these terms particularly clear.  In the NMHV sector, the first term is simply the super-attachment of the 7-field to  A(123456), and so corresponds to $H_6(123456)$. The remaining three terms all correspond to the BCFW joining of MHV sub-amplitudes. By using the Arkani-Hamed principle of analysing diagrams by repeated removal of `attached black triangles', it is easy to relate them to the 5-field ampitudes  $(67123), (67134), (67145).$ These extra terms thus correspond to $H_{5}(67123) + H_5(67134)+ H_5(67145)$, as required for $H_7(1234567)$.

The  $n$-field case in no more difficult. The total number of NMHV terms in the BCFW expansion is just $\frac{1}{2}(n-4)(n-3)$.  Again, the leading term corresponds to  the `homogeneous' term in the expansion, where the $n$th point is attached to the NMHV amplitude for $n-1$ fields. It accounts for $\frac{1}{2}(n-5)(n-4)$ of the NMHV terms. The remaining $(n-4)$ terms, by using the supersymmetric attachment rule, are easily seen to correspond one by one to the remaining simplexes in the definition of the $H_n$. Different choices of pivots simply correspond to different representations of the $H_n$, and all of these representations  are related by the `hexagon identity' which now has a fundamental geometrical meaning.

\subsection{Summary: super-amplitudes as super-volumes}

Although the details are yet to be filled in, we have a convincing specification of the NMHV super-amplitude as the super-volume of a single super-polytope, with manifest dihedral symmetry. The physical singularities correspond to momentum configurations where a vertex goes to infinity.  Spurious poles do not arise. Different representations of the super-amplitude correspond to different descriptions of the polytope. In particular, the six-term  hexagon identity expresses the (super-space equivalent of)  the four-dimensional boundary of a five-dimensional simplex  being closed. Communications from David Skinner and Lionel Mason indicate that results confirming and extending these ideas should soon be forthcoming.

\section{Outlook}

\subsection{Duality of the new integrals to twistor diagrams}

The integration performed in momentum-twistor space is essentially different from the integration over twistor space required in twistor diagrams. As an illustration of the difference, $ \langle1234\rangle$ must be non-vanishing for the new coordinates, but in the twistor diagram for any MHV process, the twistors become collinear, and so satisfy  $ \langle1234\rangle=0$. (The same collinearity arises in the twistor string model  due to Witten (2003), which has inspired so many subsequent developments.)   

Elements of the new space may be thought of {\em operators} on the primary twistor space, rather as in standard field theory, momentum arises as on operator on wave-functions $\psi(x)$. In this sense, the new integrals are dual to twistor diagrams. An amplitude $A$ is defined by equations of the form $DA = \delta$, where $D$ is a function in momentum-space, and so a differential operator in $x$-space. The principle of twistor diagrams is that they are defined for actual finite-normed wave-functions (and not for momentum states). Encoding the $\delta$-function for such states is remarkably difficult, and involves integration over many twistor variables, but once  this is achieved, it opens up a very elegant representation of $D^{-1}\delta$, giving, in principle, a completely finite amplitude for finite-normed wave-functions. This should include all the infra-red effects which arise, in this picture, even at tree-level, through the correct specification of conformal-symmetry-breaking boundaries for the twistor diagram integral.  Although twistor diagrams have, in the past, been roughly compared with Feynman diagrams, they are actually more closely analogous to integrations over $x$-space, including knowledge of the boundary of finite space-time at the null cone at infinity, which gives rise to infra-red effects.

The new integrals bear a closer analogy to Feynman diagrams, being integrals over momentum-space variables, which are left as coefficients of momentum $\delta$-functions. They offer a quite new encoding of the differential equations given by $D$. They do not {\em solve} those equations, which involve infra-red-divergent parts $A_I$ satisfying the homogeneous equation $DA_I = 0$. These solutions should still, in principle, be provided by twistor diagrams.  
 But for many purposes it is precisely the $D$, as functions of momenta, that  we want to know. Not only are they far simpler than the functionals of finite-normed wave functions, but they are closely related to the measurements of actual high-energy collision experiments. 
 
At the most fundamental level, twistor diagrams cannot be replaced by the new integrals. Another indication of their more fundamental status is that the new variables depend for their usefulness on dual conformal symmetry, which only exists within each colour-order sector, and is not defined for gravitonic scattering. (One could define the new variables with respect to any ordering, but the resulting expressions would be unlikely to demonstrate any  simplifying power.) 
Relations between amplitudes for different colour-orderings, likewise, will not have a natural expression in these co-ordinates. Yet for many purposes  the new momentum-twistor variables appear to offer a very powerful simplification. 

\subsection{Further remarks}

1. The duality between position and momentum, and its expression in twistor geometry, goes  back to the earliest ideas of Roger Penrose for the twistor programme (Penrose 1972, 1975). On one interpretation, twistors appear  as like square roots of space-time points, but on another, as square roots of particles, possessing momentum and angular momentum. This note may suggest further ways of studying such fundamentals. 

2. The principle that spurious poles  correspond to spurious boundaries should apply to the original twistor diagrams themselves. One motivation for the definition of the super-symmetric diagrams is that they become nothing but (super)-homology definitions, with a natural additive structure and the potential for spurious boundaries to disappear. But the much larger dimensionality of the integration spaces make this harder to investigate for twistor diagrams, at least beyond the MHV level.

3. The new momentum-twistor coordinates should be useful for expressing all amplitudes, by virtue of the absorption of the momentum-conservation condition that makes the cancellation of spurious poles simpler to express. Their algebraic properties alone are required for this. But it is obviously desirable that the {\em geometric} picture of the cancellation of the poles, and the emergent dihedral symmetry,  should extend beyond NMHV. One approach to achieving this would be to show that the general BCFW recursion can be represented in a geometric way, extending the special case of `attaching' a new field discussed above. This would necessarily be more complicated than the simple volume calculations discussed above, but should not go beyond purely tree-like integrals. 

Explicitly, consider the amplitudes in eight-field NNMHV sector, as originally studied in a special case by Britto, Cachazo and Feng (2004), and extended to all helicities in (Hodges 2006).

 These involve singular factors such as $[1|(2+3+4)(5+6)|7]^{-1}$, which in  momentum-twistor co-ordinates becomes $(\langle8124\rangle \langle 5678\rangle - \langle8125\rangle \langle 4678\rangle)^{-1}$. This more complicated SL4C invariant cannot arise from the simplex integrals discussed above, but it  can be given by the projective contour integral
\begin{equation}\frac{1}{(2\pi i)^{11}} \oint \frac{ DW \wedge DY \wedge DU \wedge DX }{W.Z_8\,W.Z_1\,W.Z_2 \,W.X (U.X)^2 U.Z_4\,U.Z_5 \,Y.X \,Y.Z_6\, Y.Z_7\, Y.Z_8}\, .\end{equation}
Here there is an integration over an interior $X^{\alpha}$ twistor, which makes it essentially a convolution of the simplex integrals earlier discussed. The integral is of the kind first investigated by Penrose (1972). It  has a tree structure, much simpler than the original twistor diagrams for scattering amplitudes, which even for the simplest tree-amplitude cases involve looped twistor integration. This tree structure can be generalised so as to define boundary-integrals which are the natural candidates for generalising the polytope structure beyond NMHV .

4. In particular, the BCFW recursion might profitably be considered as a rule for the merging and creation of {\em regions}. Arkani-Hamed's operations defined by the addition and removal of triangles  are essentially a special case of region creation and annihilation. 
Indeed, the concept of region may be even more useful in the twistor picture than in the usual momentum-space picture. Because the twistor diagrams are topological {\em discs}, with very striking emergent string-like properties, the concept of region naturally extends to their {\em  interiors} as well. 

5. If this picture is correct, then loop amplitudes should correspond to non-trivial integration over a loop of momentum-twistors. Contact would then be made with the many recent advances exploiting dual conformal invariant structure for such integrals.  Loop integrals in twistor space are, of course, well known: they have already been extensively studied for the representation of tree-amplitudes! Thus a mass of earlier work on the original twistor diagrams may very well take on new life in the study of loop integrals in momentum-space. If further magical identities are to be found, it may be profitable to search for them using twistor coordinates.

\section{Acknowledgements}
This work was encouraged and stimulated by participation in a workshop at the Perimeter Institute in December 2008 organised by Freddy Cachazo and Nima Arkani-Hamed, and the writer is grateful to the Perimeter Institute for its generous invitation. The expositions of the distinguished participants in the International Workshop on Gauge and String Amplitudes at Durham University, March/April 2009, especially that of Nima Arkani-Hamed, also helped to focus the questions addressed in this note. I am also grateful to my colleagues David Skinner and Lionel Mason, especially for discussion of the super-symmetric extension of the integrals.

\section{References}

N. Arkani-Hamed, F. Cachazo, and J. Kaplan, What is the simplest quantum field theory?, arXiv:0808.1446v2 (2008)
 
N. Arkani-Hamed, F. Cachazo, C. Cheung and J. Kaplan, The S-matrix in twistor space, arXiv:0903.2110v2 [hep-th] (2009)

A. Brandhuber, P. Heslop and G. Travaglini, A note on dual conformal supersymmetry of the N=4 super Yang-Mills S-matrix, arXiv:0807.4097 [hep-th] (2008)

R. Britto, F. Cachazo and B. Feng, Recursion relations for tree amplitudes of gluons,  arXiv:hep-th/0412308 (2004)

R. Britto, F. Cachazo, B. Feng and E. Witten, Direct proof of tree-level recursion relation in Yang-Mills theory,  arXiv:hep-th/0501052  (2005)

J.M. Drummond, J. Henn, V.A. Smirnov, and E. Sokatchev, Magic identities for conformal four-point integrals,
arXiv:hep-th/0607160v3 (2006)

J. M.  Drummond, J. Henn, G. P. Korchemsky and E. Sokatchev, Dual superconformal symmetry of scattering amplitudes in N=4 super-Yang-Mills theory,  arXiv:0807.1095 [hep-th] (2008)

J. M. Drummond and J. M. Henn, All tree-level amplitudes in  N=4 SYM, \newline arXiv:0808.2475v4 [hep-th] (2008)

A. Hodges, A twistor approach to the regularization of divergences, Proc. R. Soc. Lond. {\bf A 397,} 341  (1985)

A. Hodges, Twistor diagram recursion for all gauge-theoretic tree amplitudes,  hep-th/0503060  (2005a)

A. Hodges, Twistor diagrams for all tree amplitudes in gauge theory: a helicity-independent formalism, hep-th/0512336 (2005b)

A. Hodges, Scattering amplitudes for eight gauge fields, hep-th/0603101 (2006)

L. Mason and D. Skinner, Scattering amplitudes and BCFW recursion in twistor space, arXiv:0903.2083 [hep-th] (2009)

R. Penrose and M. A. H. MacCallum, Twistor theory: an approach to the quantisation of fields and  space-time, Physics Reports {\bf 4,} 241 (1972) 

R. Penrose, Twistor theory, its aims and achievements, in Quantum Gravity, eds. C. J. Isham, R.  Penrose and D. W. Sciama, Oxford University Press (1975) 

E. Witten, Perturbative gauge theory as a string theory in twistor space, arXiv:hep-th/0312171 (2003) 
 
\end{document}